\documentstyle[aps]{revtex}

\begin{document}
\title{Measureament, Trace, Information Erasure and Entropy}
\author{Qing-Yu Cai}
\address{State Key of Laboratory of Magentic Resonance and Atom and Molecular\\
Physics, Wuhan Institute of Physics and Mathematics, The Chinese Academy of\\
Sciences, Wuhan, 430071, People's Republic of China}
\maketitle

\begin{abstract}
We show that both information erasure process and trace process can be
realized by projective measurement. And a partial trace operation consists
to a projective measurement on a subsystem. We show that a nonunitary
operation will destroy the wave-behavior of a particle. We give a quantum
manifestation of Maxwell's demon and give a quantum manifestation of the
second law of therodynamics. We show that, considering the law of
memontum-energy conversation, the evolution of a closed system should be
unitary and the von Neumann entropy of the closed quantum system should be
least. 

\begin{description}
\item  PACS numbers: 03.65.-w, 89.70.+c
\end{description}
\end{abstract}

\section{INTRODUCTION}

When we attempt to apply classical mechanics and eletrodynamics to explain
atomic phenomena, they lead to results which are obviously conflict with
experiment. The mechanics which governs microword is quantum mechanics. It
is the most accurate and complete description of the world known. Thus
quantum mechanics occupies a very unusual place among physical theories: it
contains classical mechanics as a limiting case, at the same time, it
requires this limiting case for its own formulation.

Quantum mechanics tells us the evolution of a closed quantum system is
described by a unitary transformation. To observes a closed quantum system
to find out what is going on inside the system, one needs to interact the
system with an external physical system, which makes the system no longer
closed, and the unitary evolution is not necessary. Structurally, quantum
mechanics has two parts, one part concerned with quantum states, the other
with quantum dynamics. In this paper, we will show some properties about
quantum evolution.

To find out what is going on inside a quantum system, one must perform a
quantum measurement. And a partial trace operation can give the correct
description of observable quantities for subsystems of a composite system.
The relation between physical entropy and information may have been
mentioned first by Szilard [1]. Landauer's principle states that in erasing
one bit of information, on average, at least $k_{B}T\ln (2)$ energy is
dissipated into the environment ( where $k_{B}$ is Boltzmann's constant and $%
T$ is the temperature of the environment at which one erases.) [2].
Piechocinska derived Landauer's principle from microscopic considerations
[3].

In this paper, we first give a definition of the wave behavior of a quantum
system. Suppose a quantum system is in state 
\begin{equation}
|\varphi >=\sum_{i}a_{i}|i>,
\end{equation}
where $\sum_{i}|a_{i}|^{2}=1$, $\{|i>\}$ is a set of orthogonal vectors. If $%
i\geq 2$, we will say that this quantum system has wave behavior. Else,
there is only particle behavior in this quantum system. The quantum
operations formalism is a general tool for describing the evolution process
of quantum systems, such process include unitary evolution, quantum
measurement, and even more general processes [4]. A general quantum
dynamical process is described by a quantum operation. We will use quantum
operations formalism to described the quantum evolution processes in this
paper. We use density operators formalism $\rho $ or Dirac's bracket
notation $|\psi >$ to described the state of quantum system. In this paper,
we will use a word: momentum-energy conservation. The momentum-energy
conservation is a word for short. In microword, the conversation quantities
would include momentum, energy, spin and other quantity. All these
quantities we call them momentum-energy conservation in this paper. The
entropy of a classical system is defaulted as Shannon entropy. And the
entropy of a quantum system is considered as the von Neumann entropy default
in this paper.

\subsection{Trace}

The trace of a matrix A is defined to be the sum of its diagonal elements, 
\begin{equation}
tr(A)\equiv \sum_{i}A_{ii}.
\end{equation}
The trace is cyclic and linear. Suppose we have physical systems A and B,
whose state is described by a density operator $\rho ^{AB}$. The reduced
density operator for system A is defined by 
\begin{equation}
\rho ^{A}\equiv tr_{B}(\rho ^{AB}),
\end{equation}
where tr$_{B}$ is a map of operators known as the partial trace over system
B. The partial trace is defined by 
\begin{equation}
tr_{B}(|a_{1}><a_{1}|\otimes |b_{1}><b_{1}|)\equiv
|a_{1}><a_{1}|tr(|b_{1}><b_{1}|).
\end{equation}
The partial trace operator is the operation which gives description of
observable quantities for subsystems of a composite system.

\subsection{Measurement}

Distinguishing quantum states needs quantum measurement. Quantum
measurements are described by a collection $\left\{ M_{m}\right\} $ of
measurement operators [5]. These are operators acting on the state space of
the system being measured. The index $m$ refers to the measurement outcomes
that may occur in the experiment, 
\begin{equation}
p(m)=<\psi |M_{m}^{\dagger }M_{m}|\psi >,
\end{equation}
where $|\psi >$ is the state of the quantum system immediately before the
measurement. The state of the system after the measurement is 
\begin{equation}
\frac{M_{m}|\psi >}{\sqrt{<\psi |M_{m}^{\dagger }M_{m}|\psi >}}.
\end{equation}
And the completeness equation expresses the fact that probabilities sum to
one: 
\begin{equation}
1=\sum_{m}p(m)=\sum_{m}<\psi |M_{m}^{\dagger }M_{m}|\psi >.
\end{equation}
The quantum measurements postulate gives a rule describing the measurement
statistics and a rule describing the postmeasurement state of the system. It
is familiar that quantum mechanics describe $projective$ $measurements$,
`Positive Operator-Valued Measure' measurements ($POVMs$), and a $general$ $%
measurements$. A projective measurement is described by a Hermitian
operator, $M$, on the state space of the system being observed 
\begin{equation}
M=\sum_{m}mP_{m},
\end{equation}
where $P_{m}$ is the projector onto the eigenspace of $M$ with eigenvalue $m$%
. The possible outcomes of measurement correspond to the eigenvalues, $m$,
of the observable. Upon measuring the state $|\psi >$, the probability of
getting result $m$ is given by 
\begin{equation}
p(m)=<\psi |P_{m}|\psi >.
\end{equation}

\subsection{Information Erasure}

The bit is a fundamental unit of information. We will assume to have a large
number of bits but they will be erased individually, one by one. Landauer
argues that since information erasure is a logical function which does not
have a single-valued inverse, it must be associated with physical
irreversibility and require heat dissipation. Suppose we have a quantum
system which is in the unknown state $\rho $. A information erasure process
is that we prepare this system in a standard state, 
\begin{equation}
\epsilon :\rho \rightarrow |0><0|,
\end{equation}
where $\rho $ is an arbitrary state and $|0><0|$ is a standard state. This
information erasure process is nonunitary generally. If the system is in a
known state, this information erasure process can be realized by a unitary
operation. But in the case we have a large number of bits to erase, the
information erasure operation must be irreversible. In this paper,
information erasure will be considered as the quantum state erasure of a
quantum system.

\section{Relation between Measurement, Trace, and Erasure}

Let $H_{A}$ be any Hilbert space, spanned by an orthonormal basis $|1>$ $%
...|d>$. Then the trace map $\rho \rightarrow tr(\rho )$ can be represented
as 
\begin{equation}
tr(\rho )=\sum_{i=1}^{d}<i|\rho |i>.
\end{equation}
If $H_{A}$ is both the input and output space, a quantum measurement can be
described: 
\begin{equation}
M_{m}(\rho )=\sum_{i=1}^{d}|m><i|\rho |i><m|,
\end{equation}
where $M_{m}$ is a quantum measurement operator.

Let $H_{Q}$ be any input Hilbert space, spanned by an orthonormal basis $|1>$
$...|d>$, and let $H_{Q}^{\prime }$ be a one dimensional output space,
spanned by the state $|0>$. Then, information erasure can be described like
this 
\begin{equation}
R(\rho )=\sum_{i=1}^{d}|0><i|\rho |i><0|.
\end{equation}
where $R$ is an erasure operator. No matter what the state $\rho $ is, the
state of output space is $|0><0|$.

A trace operation can be realized by a projective measurement when we never
learn the result of the measurement. Suppose the a quantum in the state $%
|\psi >$, the state after the measurement is given by 
\begin{equation}
\rho ^{\prime }=\sum_{i}M_{i}|\psi ><\psi |M_{i}^{\dagger }.
\end{equation}
Relation between trace operation and the measurement can be described like
this, 
\begin{equation}
1=tr(|\psi ><\psi |)=\sum_{i}<\psi |M_{i}^{\dagger }M_{i}|\psi >.
\end{equation}
Suppose we have a joint system $AQ$ in the state 
\begin{equation}
\rho ^{AQ}=\sum_{i,j}p_{ij}\rho _{i}^{A}\otimes |j_{Q}><j_{Q}|,
\end{equation}
and wish to trace out the subsystem $Q$. Then the state of the subsystem $A$
becomes 
\begin{equation}
\rho ^{A}=tr_{Q}(\rho ^{AQ})=\sum_{j^{\prime }}<j_{Q}|\rho
^{AQ}|j_{Q}>=\sum_{i,jj^{\prime }}p_{ij}\rho _{i}^{A}\delta _{j,j^{\prime }}.
\end{equation}
Suppose that a projective measurement described by $P_{j}^{Q}$ is performed
on the quantum system $Q$, but we $never$ $learn$ the result of the
measurement. The state of the system $A$ after a projective measurement on
the system $Q$ is given by 
\begin{equation}
\rho ^{A}=\sum_{j^{\prime }}P_{j}^{Q}\rho ^{AQ}P_{j}^{Q}=\sum_{i,jj^{\prime
}}p_{ij}\rho _{i}^{A}\delta _{j,j^{\prime }}.
\end{equation}
We can see that a project measurement on a subsystem is the same as a
partial trace. To see it clearly, let us consider a GHZ state [6] 
\begin{equation}
|GHZ>=\frac{1}{\sqrt{2}}(|0_{A}0_{B}0_{C}>+|1_{A}1_{B}1_{C}>)
\end{equation}
of a three qubit system $ABC$. Then trace one qubit (suppose of system $A$)
out of the three qubit system. It has 
\begin{equation}
tr_{A}(|GHZ><GHZ|)=\frac{1}{2}\rho _{00}^{BC}+\frac{1}{2}\rho _{11}^{BC},
\end{equation}
where $\rho _{00}^{BC}=|0_{B}0_{C}><0_{B}0_{C}|$, $\rho
_{11}^{BC}=|1_{B}1_{C}><1_{B}1_{C}|$. From Eq.(20), we know that after a
partial trace on system $A$, the system of $BC$ has a probability $p=\frac{1%
}{2}$ in state $\rho _{00}^{BC}$ or in the state $\rho _{11}^{BC}$.
Obviously, the quantum measurement postulate tells us that is we perform a
project measurement in basis $\left\{ |0>,\text{ }|1>\right\} $ on the
system $A$, the measurement result is in state $|0>$ ( consists to $\rho
_{00}^{BC}$ ) or in state $|1>$ ( consists to $\rho _{11}^{BC}$ ) with a
probability $p=\frac{1}{2}$. Consider the case the three qubit in the state $%
|w>$ [7] 
\begin{equation}
|w>=\frac{1}{\sqrt{3}}%
(|1_{A}0_{B}0_{C}>+|0_{A}1_{B}0_{C}>+|0_{A}0_{B}1_{C}>).
\end{equation}
Let us suppose we trace out the qubit of system $A$. then we have 
\begin{equation}
tr_{A}(|w><w|)=\frac{1}{3}\rho _{00}^{BC}+\frac{2}{3}|\psi _{BC}^{+}><\psi
_{BC}^{+}|,
\end{equation}
where $|\psi _{BC}^{+}>=\frac{1}{\sqrt{2}}(|0_{B}1_{C}>+|1_{B}0_{C}>)$ is a
Bell state. Clearly, after a project measurement on system $A$ in basis $%
\left\{ |0>,\text{ }|1>\right\} $, the system $A$ has a probability $p=\frac{%
1}{3}$ in the state $|1>$ (consists to $\rho _{00}^{BC}$ ) and a probability 
$p=\frac{2}{3}$ in the state $|0>$ ( consists to $|\psi _{BC}^{+}>$ ).

Information erasure operation can be realized by a quantum projective
measurement and a unitary operation. First, we perform a projective
measurement. Then the quantum system will be in a known state (suppose in
the state $|j><j|$) after the quantum measurement. We can perform a unitary
operation on the state $|j><j|$ to prepare the system in state $|0><0|$.
Suppose we want to erase the state information of quantum system $A$, which
is in an known state $\rho $. First, we can perform a projective measurement
described by projectors $P_{j}$. When we gain the measurement result
(suppose in state $|j>$), then we can perform a unitary operation $U$ to
prepare the system in a state $|0>$, 
\begin{eqnarray}
M &:&\rho \rightarrow |j><j|, \\
U &:&|j><j|\rightarrow |0><0|.
\end{eqnarray}
This is a information erasure process.

\section{Wave-particle duality:unitary and nonunitary operations}

The evolution of the state of a closed quantum system is described by the $%
Schr\ddot{o}dinger^{\prime }s$ equation. A general quantum dynamical process
is described by a quantum operation. Complementarity principle tells us the
microscopic word has the behavior, wave-particle duality. One can not draw
pictures of individual quantum processes [8]. To gain information of a
quantum system, one has to perform a quantum measurement. A unitary
operation on a quantum system will keep the wave behavior of this system.
But, a nonunitary operation, will destroy the wave behavior of the system.

\subsection{Nonunitary operations and particle behavior}

In the quantum case, the completeness relation requires that trace of $\rho $
equal to one, $tr(\rho )=1$. We can see it consists to the quantum
measurement, Eq.(7). A trace process can be treated as a notion that we can
find out the particle in the whole space to a certain. After a trace process
on a quantum system, the wave behavior disappeared completely. If the
quantum system is $trace-preserving$ ( $tr(\rho )=1$ ), which means that the
trace of this quantum system is unit, the probability of find out the
particle is 1.

Consider the case of quantum measurement. When we perform a projective
measurement on a quantum system, the result of the measurement is in a state
with the probability give by Eq.(5). Since the state is in an orthogonal
state after a projective measurement, according to the definition of Eq.
(1), the behavior of this system is particle. The wave behavior disappeared.
But, $POVMs$ maybe show some wave behavior. Suppose a particle is a two
dimensions $\left\{ |0>,|1>\right\} $ quantum system. Consider a $POVM$
containing three elements [5], 
\begin{equation}
E_{1}=\frac{\sqrt{2}}{1+\sqrt{2}}|1><1|,
\end{equation}
\begin{equation}
E_{2}=\frac{\sqrt{2}}{1+\sqrt{2}}\frac{(|0>-|1>)(<0|-<1|)}{2}
\end{equation}
\begin{equation}
E_{3}=I-E_{1}-E_{2}
\end{equation}
We can see that $E_{1}$, $E_{2}$, and $E_{3}$ are not orthogonal to each
other. Then after a POVM, the quantum system will keep some wave behavior
and appear some particle behavior.

A information erasure operation will destroy the wave behavior completely.
After a information erasure operation, the system is in the state $|0><0|$.
There is no wave behavior in this quantum system. Since the information
erasure operation is physical irreversible, the wave behavior of the quantum
system can not be recurred.

\subsection{Quantum operation and von Neumann entropy}

Entropy is a key concept of quantum information theory. It measures how much
uncertainty there is in the state of a physical system. Von Neumann defined
the entropy of a quantum state by the formula [9] 
\begin{equation}
S(\rho )\equiv -tr(\rho \log \rho ).
\end{equation}
If $\lambda _{x}$ are the eigenvalues of $\rho $ then von Neumann's
definition can be re-expressed 
\begin{equation}
S(\rho )=-\sum_{x}\lambda _{x}\log \lambda _{x},
\end{equation}
where $0log0\equiv 0$. The entropy of a system measures the amount of
uncertainty about the system before we learn its value. It is a measure of
the ``amount of chaos'' or of the lack of information about a system. If one
has complete information, i.e., if one is concerned with a pure state, $%
S(\rho )=0$ [10].

How does the entropy of a quantum system behave when we perform an operation
on that system? Obviously, a unitary operation does not change the entropy
of a system because a unitary transformation does not change the eigenvalues
of $\rho $. In general, a nonunitary transformation would change the
eigenvalues of $\rho $. So a nonunitary operation would change the von
Neumann entropy of a quantum system.

Suppose $P_{i}$ is a complete set of orthogonal projectors and $\rho $ is a
density operator. Then the entropy of the state $\rho ^{\prime
}=\sum_{i}P_{i}\rho P_{i}$ of the system after the measurement is at least
as great as the original entropy, 
\begin{equation}
S(\rho ^{\prime })\geq S(\rho ),
\end{equation}
with equality if and only if $\rho =\rho ^{\prime }$ [5]. From Eq.(9), we
know the measurement outcomes $P_{m}$ is gained randomly with the
probability $m$. So the uncertainty of the system increases after under the
projective measurement if we never learn the result of the measurement. The
entropy of the system would increase under a projective measurement. But how
does the entropy behave depends on the type of measurement which we perform.
A projective measurement increases entropy of a quantum system. But, a
generalized measurements can decrease entropy of a quantum system.

Consider a trace operation. The information of a quantum system disappeared
completely after a trace operation. So the uncertainty of the quantum system
would increase. The uncertainty of the system would increase. In another
view, a trace process can be realized by a projective measurement if we do
not know the measurement result. So a trace operation would increase the
entropy of the system.

Information erasure process will induce the entropy of the environment
increase. Suppose a quantum system is in a state $\rho $. The von Neumann
entropy of the system is 
\begin{equation}
S(\rho )\geq 0,
\end{equation}
where $S(\rho )=0$ if and only if $\rho $ is a pure state. After the
information erasure operation, the system is in the known state $|0><0|$.
The entropy of the system is zero. The entropy of the system is
nonincreasing. Since entropy does not decrease, the entropy of the
environment must increase. To realize the information erasure process, there
must be interaction between system and environment. A information erasure
process has to exchange momentum-energy with environment randomly. In
another view, a information erasure process can be realized by a projective
measurement and a unitary operation. A unitary operation does not change the
entropy of the system. The change of entropy is derived by the measurement.
A measurement will induce exchange of momentum-energy between system and
environment. Here, the projective measurement will decrease the entropy of a
system because we have to know the result of the measurement if we want to
realize the information erasure operation.

\section{Quantum measurement and the second law of thermodynamics}

The second law of thermodynamics states that the entropy in a closed system
can never decrease. In 1871, J. C. Maxwell proposed the existence of a
machine that apparently violated this law. Suppose we have a vessel in which
full of air at unitary temperature are moving with velocities by no means
unitary. And such a vessel is divided into two portions, A and B, by a
division in which there is a small hole. Let suppose that a demon, who can
see the individual molecules, opens and closes this hole, so as to allow
only the swifter molecules to pass from A to B, and only the slow from B to
A. He will thus, without expenditure of the work, raise the temperature of B
and lower that of A, in contraction to the second law of thermodynamics [11].

The resolution to the Maxwell's demon paradox lies in the fact that the
demon must perform measurement on the molecules moving between the
partitions, in order to determine their velocities. The result of this
measurement must be stored in the domon's memory. Because any physical
memory is finite, the demon must erase information from its memory, in order
to have space for new measurement results. By Landauer's principle, erasing
information increases the total entropy of the combined system which
includes demon, gas vessel, and their environments. The second law of
thermodynamics is obeyed.

We give a quantum manifestation of Maxwell's demon. The demon and molecules
in the vessel are all considered as quantum systems. To determine the
velocities of molecules, the demon has to perform a quantum projective
measurement. The average velocities of the molecules is 
\begin{eqnarray}
\overline{M} &=&\sum_{m}mP(m)  \nonumber \\
&=&<\psi |M|\psi >,
\end{eqnarray}
where $M$ is a velocity operator. The standard deviation is 
\begin{eqnarray}
\triangle (M) &=&<(M-<M>)^{2}>  \nonumber \\
&=&<M^{2}>-<M>^{2}.
\end{eqnarray}
It is a measure of the typical spread of the observed values upon
measurement of $M$. In the vessel, the spread of the velocity of molecules
is a Boltzmann's spread. So the standard deviation is nonzero. Every demon's
measurement on the molecules will exchange some of momentum-energy between
demon and molecules. We can calculate the least value of the momentum
exchange under every measurement. Suppose the volume of the vessel is $L^{3}$%
. From the Heisenberg uncertainty principle, we know that 
\begin{equation}
\triangle x\triangle p\sim 
\rlap{\protect\rule[1.1ex]{.325em}{.1ex}}h%
.
\end{equation}
The uncertainty of momentum is at least $
\rlap{\protect\rule[1.1ex]{.325em}{.1ex}}h%
/L$ in every measurement in one axes. Every time, when demon want to
determine the velocity of molecules, there is some uncertainty momentum, at
least $
\rlap{\protect\rule[1.1ex]{.325em}{.1ex}}h%
/L$, exchanged between demon and molecules. The system of vessel is not a
closed. So the entropy of the vessel can decrease. Considering the whole
system including molecules, demon, and environment, the entropy of combined
system does not decrease.

Let us see a laser cooling of free atoms case. The general method of cooling
is to reduce the kinetic energy of an atom after it was loaded into the
trap. Assume we have an unbound gas of atoms. After a laser cooling process,
the atoms can be in a Bose-Einstein condensation state [12]. The entropy of
the atoms decreases all long in this laser cooling process. But the entropy
of the whole system including atoms and photon field increases. This laser
cooling process is very like the maxwell's demon paradox.

Consider a closed system including a large number of molecules. The average
velocity of molecules is $\overline{v}$. Every molecule has a nonzero
velocity in this system. Every molecule collides with others randomly all
the time. There is momentum-energy exchange under every collision. The
collision is a irreversible process in natural state. In a closed system, a
collision process is nonreversible. Since every collision is random, the
chaos of the system would increase after a long time. The entropy of this
closed system increases.

\section{the principle of least action in microword}

There is only one physical process in reality. It is well known there is the
principle of least action in the classical word. It corresponds to the
Fermat's principle in optics. The principle of least action tells us that
every one in classical word will has a certain position and momentum
forever. There is only one path is realizable in every possible path.

Quantum evolution of a closed system is a unitary evolution which can be
described by Schr\"{o}dinger's equation. A state of a closed system is
determined. As we have known, microword is a linear word. The linearity of
the microword can be considered as the requirement of momentum-energy
conservation and quantization. Suppose we have a two-level atom in the state 
\begin{equation}
|\Phi >=\alpha |0>+\beta |1>,
\end{equation}
where $|0>$ is the ground state and $|1>$ is the excited state, $|\alpha
|^{2}+|\beta |^{2}=1$. The energy level of the ground state is $E_{0}$, and
the excited state is $E_{1}$. Suppose there is a photon with energy $%
\triangle E=\frac{1}{2}(E_{0}-E_{1})$ which was absorbed into the system. If
the system is in the state $|0>$ at first, then it should be in the state 
\begin{equation}
|\Phi >=\frac{1}{\sqrt{2}}(|0>+|1>)
\end{equation}
after it absorbed the photon. But a classical probability view is that this
system is either in state $|0>$ or in state $|1>$ with certain. It is easy
to see that a classical probability view will induce the energy not
conservation. If the system is in state $|1>$ after absorbed a photon with
energy $\triangle E$, there is another $\triangle E$ required of this
combined system. Then the energy of this combined system is not conservation
any longer. If the system is still in the state $|0>$ after it absorbed this
photon, there is some energy $\triangle E$ unwanted. The energy of the
combined system is not conservation.

Another example is the double-slit interference. Reasonably, we can suppose
the transverse momentum of the particle is zero. The state of particles
exiting an interferometer is 
\begin{equation}
|\Psi >=\frac{1}{\sqrt{2}}[|\psi _{1}(r)>+|\psi _{2}(r)>],
\end{equation}
where $|\psi _{1}(r)>$ and $|\psi _{2}(r)>$ represent the possibility for
the particles to take path 1 or 2. If the particle only go through path 1 or
path 2 every time, the transverse momentum is not conservation yet.

Comparing to the principle of least action in classical, we can see that
entropy is least in the microword. The von Neumann entropy is least of a
closed system, which give a limitation that the evolution of a closed system
is unitary and the a microword is linear. In this case, we can say that the
least of von Neumann entropy is an elementary requirement in the microword
comparing to the principle of least action in classical word if and only if
the momentum-conservation of energy and the quantization is needed in the
microword. There is only one physics in reality. In classical word, it can
be manifestation by the principle of least action. In the microword, it can
be described as that the von Neumann entropy of a closed system is least.

\section{conclusion and suggestions for future research}

We have presented that both a trace operation and a information erasure
process can be realized by a projective measurement in physics. A partial
trace can be realized by a projective measurement on a subsystem if we never
know the result of the measurement. We have discussed the relation between
quantum operations and the wave-particle duality behavior in microword. We
have given a quantum manifestation of Maxwell's demon. We have shown that
the quantum measurement would increase the entropy of the environment. We
have given a manifestation of the principle of least action in microword.
And we have presented that the quantization and the conservation of
momentum-energy requires the evolution of a closed system is a unitary and
the entropy of this system is least.

The information erasure process is very important in quantum computation.
The memory of the quantum computer has to be erased before a new job started
since deleting an unknown state is impossible [13]. The relation between
information erasure and heat dissipation in a quantum system should be
studied deeply in the future. And the quantum manifestation of the Maxwell's
demon maybe be practical to calculate the relation between decrease of
entropy of the vessel and the exchange of energy between demon and molecules.

\section{acknowledgments}

The author acknowledge financial support from the National natrual Science
Foundation of China ( Grant No 10004013 ).

\section{REFERENCES}

[1] L. Szilard, Z. Phys. 53, 840 (1929).

[2] R. Landauer, IBM J. Res. Dev. 3, 183 (1961).

[3] B. Piechocinska, Phys. Rev. A 61, 062314

[4] M. A. Nielsen, C. M. Dawson, J. L. Dodd, A. Gilchrist, D. Mortimer, T.
J. Osborne, M. J. Bremner, A. W. Harrow, and A. Hines, Phys. Rev. A 67,
052310 (2003)

[5] M. A. Nielsen and I. L. Chuang, Quantum Computation and Quantum
Information (Cambridge University Press, Cambridge, UK, 2000).

[6] D. M. Green, M. Horne, and A. Zeilinger, in Bell's Theorem, Quantum
Theory, and Conceptions of the Universe, edited by M. Kafatos (Kluwer,
Dordrecht, 1989); D. M. Greenberger, M. A. Horne, A. Shimony, and A.
Zeilinger, Am. J. Phys. 58, 1131 (1990); N. D. Mermin, Phys. Today 43 (6), 9
(1990).

[7] A. Zeilinger, m. A. Horne, and D. M. Greenberger, NASA Conf. Publ. No.
3135 (National Aeronautics and Space Adiministration, Code NTT, Washington,
DC, 1997).

[8] N. Gisin, G. Ribordy, W. Tittel, and H. Zbinden, Rev. Mod. Phys. 74, 145
(2002).

[9] von Neumann, G\={o}tt. Nachr. 273; A. Einstein, 1914, Verh. Dtsch. Phys.
Ges. 12, 820.

[10] A. Wehrl, Rev. Mod. Phys. 50 (2), 221 (1978).

[11] J. C. Maxwell, Theory of Heat. Longmans, Green, and Co., London, 1871;
L. Szilard. Z. Phys., 53, 840 (1929); C. H. Bennet, Int. J. Theor. Phys.,
21, 905 (1982); C. H. Bennett, Sci. Am., 295(5), 108 (1987).

[12] D. J. Wineland and Wayne M. Itano, Phys. Rev. A 20, 1521 (1979);
Anthony J. Leggett, Rev Mod. Phys., 73, 307 (2001).

[13] Arun Kumar Pati and Samuel L. Braunstein, 404, 164 (2000).

\end{document}